\documentclass[12pt]{article}
\usepackage{epsfig}
\usepackage{url}
\makeatletter
\def\alt{\mathrel{\mathpalette\gl@align<}}
\def\agt{\mathrel{\mathpalette\gl@align>}}
\def\gl@align#1#2{\lower.6ex\vbox{\baselineskip\z@skip\lineskip\z@
\ialign{$\m@th#1\hfil##\hfil$\crcr#2\crcr\sim\crcr}}}
\makeatother

\begin{document}
\begin{flushright}
MIFPA-11-30\\
July, 2011
\end{flushright}
\vspace*{1.0cm}
\begin{center}
\baselineskip 20pt
{\Large\bf
$B_s \to \mu^+\mu^-$
in Supersymmetric Grand Unified Theories
} \vspace{1cm}

{\large
Bhaskar Dutta$^*$, Yukihiro Mimura$^\#$ and Yudi Santoso}
\vspace{.5cm}

$^*${\it
Department of Physics, Texas A\&M University,
College Station, TX 77843-4242, USA
}\\
$^\#${\it
Department of Physics, National Taiwan University, Taipei,
Taiwan 10617, R.O.C.
}\\
\vspace{.5cm}

{\bf Abstract}
\end{center}

We investigate the recent CDF measurement of the Br$(B_s \to \mu^+\mu^-)$ 
which shows excess over the Standard Model. We consider minimal supergravity 
motivated models (mSUGRA)/CMSSM and grand unified  models, SU(5) and SO(10).  
In the grand unified models, the neutrino mixings provide an additional source 
of squark flavor violation through the quark-lepton unification. 
In the context of minimal SU(5) model, we find that  the new CDF measurement 
has imposed a lower bound on  the branching ratio of $\tau\rightarrow\mu\gamma$ 
for a large CP phase in the $B_s$-$\bar B_s$ mixing. 
Recall that there have been indication for a large CP phase in $B_s$ mixing 
from $B_s\rightarrow J/\psi\phi$ (Tevatron and LHCb) and dimuon asymmetry (D0). 
We also predict  Br($\tau\rightarrow\mu\eta$) for the possible range of values of 
Br($\tau\rightarrow\mu\gamma$).

\thispagestyle{empty}

\bigskip
\newpage

\addtocounter{page}{-1}

\section{Introduction}
\baselineskip 18pt

Recently, the CDF collaboration has reported
the measurement of the 
rare decay process of $B_s$ meson to $\mu^+\mu^-$ \cite{CDF},
and the branching ratio of the decay is  
\begin{equation}
{\rm Br} (B_s \to \mu^+\mu^-)^{\rm CDF} = (1.8^{+1.1}_{-0.9}) \times 10^{-8}.
\label{1sigma}
\end{equation}
The 90\% confidence level (CL) range is
\begin{equation}
4.6 \times 10^{-9} < {\rm Br} (B_s \to \mu^+\mu^-)^{\rm CDF} < 3.9 \times 10^{-8}.
\label{90CL}
\end{equation}
%
The measured branching ratio at the 90\% CL is deviated from
the standard model (SM) prediction \cite{Buchalla:1993bv,Gamiz:2009ku}:
\begin{equation}
{\rm Br} (B_s \to \mu^+\mu^-)^{\rm SM} 
= (3.19 \pm 0.19) \times 10^{-9}.
\end{equation}
The SM prediction above is calculated from the ratio of
the branching ratios and the $B_s$ mass difference, $\Delta M_s$.
The discrepancy cannot be explained by the
hadronic uncertainties,
which comes from the uncertainty on the bag parameter $B_{B_s} = 1.33 \pm 0.06$ \cite{Gamiz:2009ku}.
Thus, this discrepancy implies the existence of a new physics (NP)
beyond the standard model.
The excess can be soon verified at the LHC~\cite{Aaij:2011rj}.

The excess of the $B_s\to \mu^+\mu^-$ rare decay
can be reproduced in many NP models.
Supersymmetry (SUSY)
is one of the promising candidates for the new physics.
In the minimal SUSY extension of the standard model (MSSM),
the rare decay of $B_s \to \mu^+\mu^-$
is induced by the neutral Higgs mediated flavor changing
operator \cite{Choudhury:1998ze}.
The operator is proportional to $\tan^3\beta/m_A^2$,
where $\tan\beta$ is the ratio of vacuum expectation values
of the up-type and down-type Higgs fields,
and $m_A$ is the mass of the CP odd Higgs field.
Therefore, the branching ratio is proportional to $\tan^6\beta$,
and the excess of this rare decay process
implies a large value of $\tan\beta$ 
for the experimentally allowed parameter region of the model,
where the masses of SUSY particles (especially for stops) are 
bounded from below.
The flavor changing Higgs coupling is induced by
finite corrections in the down-type quark mass matrix,
and it can be generated even in the minimal flavor violating
models where the CKM (Cabibbo-Kobayashi-Maskawa) quark mixing 
is the only source of the flavor violation.
In this case, the operator is generated by a chargino-stop loop.

In the SUSY grand unified theories (GUTs),
flavor mixing in the lepton sector (i.e. neutrino mixing)
can cause a squark flavor violation
due to the quark and lepton unification \cite{Borzumati:1986qx,Barbieri:1994pv}.
The squark flavor violation can induce an additional contribution
to the Higgs mediated flavor changing coupling via a gluino-sbottom loop.
In the minimal type of SU(5) GUT,
the right-handed down-type quark and the left-handed lepton doublet
are unified in a multiplet,
the right-handed down-type squarks have the flavor violating source \cite{Moroi:2000tk},
and the left-handed squarks (i.e. both up- and down-type)
can have the flavor violating sources in SO(10) GUTs,
where the size of the squark/slepton flavor violation can be related to
the enhancement of the proton life time~\cite{Dutta:2007ai}.
The detail investigation of the flavor violating processes
are important to find the footprint of the 
SUSY GUTs \cite{Parry:2005fp,Dutta:2006gq,Goto:2007ee,Hisano:2008df}.

The right-handed squark flavor violation
with a large $\tan\beta$
can induce a sizable $B_s$-$\bar B_s$ mixing amplitude
by a flavor changing Higgs mediation
rather than the box diagram 
contribution \cite{Hamzaoui:1998nu,Buras:2001mb,Foster:2004vp,Dutta:2009iy}.
The $b$-$s$ flavor violation, 
which is suggested by $B$ decay experiments
(the CP violation in $B_s \to J/\psi \phi$ decay \cite{Aaltonen:2007he, LHCb-phis}
and dimuon asymmetry in semileptonic $B$ decays \cite{Abazov:2010hv}),
%
can be related to 
the $\tau$-$\mu$ flavor violation, such as
$\tau\to\mu\gamma$, $\tau\to 3\mu$ and $\tau\to \mu\eta$,
in SUSY GUTs.
%
%
Due to the bounds of the flavor violating $\tau$ decays,
especially for $\tau\to\mu\gamma$,
the size of the $b$-$s$ flavor violation
and the CP odd Higgs mass
can be bounded to obtain a large CP phase.
As a result,
the existence of a large CP phase
for $B_s$-$\bar B_s$ mixing 
can constrain Br($B_s\to \mu^+\mu^-$)
in the case where the quark and lepton unification is manifested.
In fact, we obtain
that the Br($B_s \to \mu^+\mu^-$)
should be larger than $O(10^{-8})$ \cite{Dutta:2009hj}.
Inversely speaking,
the branching ratio measured by CDF
gives a bound on the flavor violating $\tau$ decays.
In the paper,
we will investigate the bound of $\tau$ decays
in SUSY GUTs.

This paper is organized as follows: in section II, we discuss  $B_s\to \mu^+\mu^-$ 
in minimal supergravity (mSUGRA), in section III, we discuss large $B_s$ mixing phase 
in SUSY GUT models, in section IV, we discuss the flavor violating $\tau$ decays 
and we conclude in section V.


\section{$B_s\to \mu^+\mu^-$ in minimal supergravity (mSUGRA)}

The rare decay of $B_s\to\mu^+\mu^-$
can be generated
even if the CKM quark mixing is the only source of
flavor violation in MSSM.
We first study the parameter space
in minimal supergravity model (mSUGRA) (constrained MSSM (CMSSM))~\cite{msugra}
to realize the CDF measured $B_s\to\mu^+\mu^-$.

In mSUGRA,
the parameters are the universal scalar mass ($m_0$),
the unified gaugino mass ($m_{1/2}$),
the universal trilinear scalar coupling ($A_0$),
the ratio of the Higgs vev ($\tan\beta$), and the sign of the Higgs mixing parameter $\mu$.
The current experimental constraints,
$b\to s\gamma$ and the lightest Higgs mass bounds
raise the squark masses. 
As a result,
lower $\tan\beta$ is disfavored
to obtain Br($B_s\to\mu^+\mu^- ) \agt 10^{-8}$~\cite{Arnowitt:2002cq}. 

Recent results from ATLAS with 165 pb$^{-1}$ of luminosity show that  
the maximum ruled out  value of $m_{1/2}$ is $\sim 450$ GeV \cite{ppc}, 
which corresponds to squark, gluino masses $\sim$ 1 TeV.

\begin{figure}[tbp]
 \center
  \includegraphics[width=8cm]{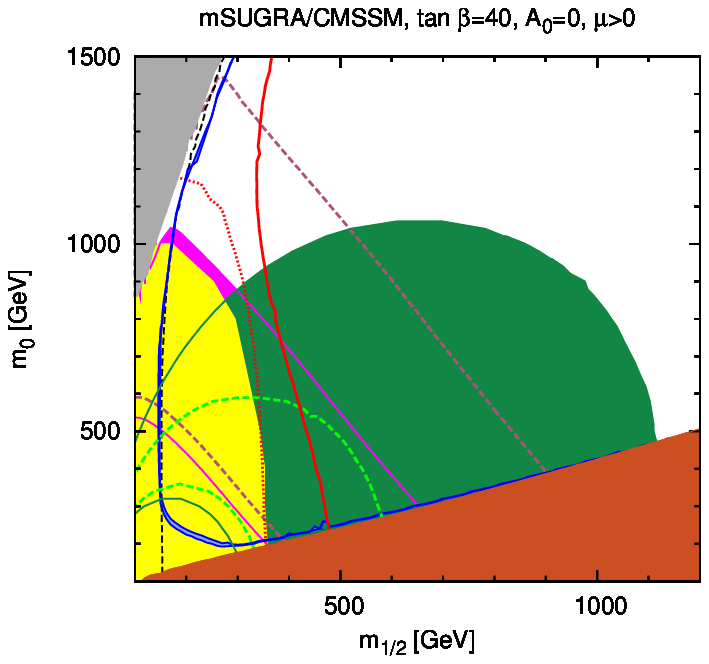}
  \includegraphics[width=8cm]{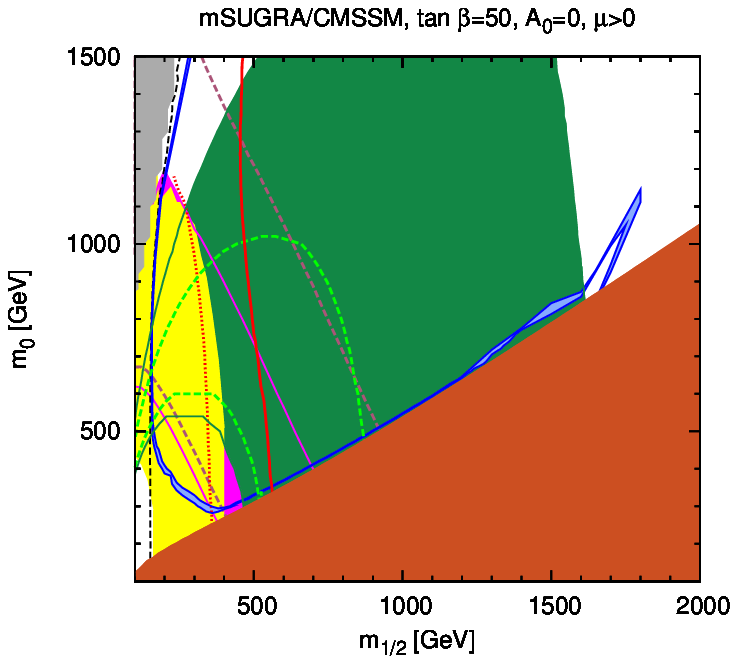}
 \caption{The mSUGRA/CMSSM parameter space is shown. 
The CDF allowed range of BR($B_s\to\mu^+\mu^-$) is shown by the green shaded region. 
The other shaded regions and lines are described in the text.
}
\label{Fig1}
\end{figure}

In Fig.\ref{Fig1}, we show a typical parameter space for mSUGRA, with $\tan\beta=$40 and 50. 
The model parameters
are already significantly constrained by different experimental results. 
The most important constraints for limiting the parameter space are: 
(i)~the light Higgs mass bound of $m_{h^0} > 114.4$~GeV from LEP~\cite{higgs1} 
(red dotted line shows the contours of $m_{h^0} = 114.4$~GeV, 
as calculated using {\tt FeynHiggs-2.6.5}~\cite{FeynHiggs});
(ii)~the $b\rightarrow s \gamma$ branching ratio~\cite{bsgamma}
(95\% CL excluded in the yellow shaded region in Fig.~\ref{Fig1}); 
(iii)~the 2$\sigma$ bound on the dark matter
relic density: $0.106 < \Omega_{\rm CDM} h^2 <0.121$ from WMAP~\cite{wmap} 
(blue region in Fig.~\ref{Fig1});
(iv)~the bound on the lightest chargino mass of $m_{\tilde\chi^\pm_1} >$
103.5~GeV from LEP \cite{aleph} (region left to the black dashed line is excluded) 
and (v) the muon magnetic moment anomaly
$a_\mu = (g_\mu-2)/2$ (pink shaded region in Fig.~\ref{Fig1} is within 2$\sigma$ 
of where one gets a 4$\sigma$ deviation from the SM as suggested by the
experimental results~\cite{amu} and the analysis in~\cite{Teubner:2010ah}). 
We also show the 2$\sigma$ contours from 3.2$\sigma$ deviation based on~\cite{Davier:2009zi} 
by slanted dashed purple lines. 
These two references use recent changes in the hadronic contribution to 
calculate the leading order hadronic contribution. Assuming that the future data
confirms the $a_{\mu}$ anomaly, the combined effects of $g_\mu -2$ and
$m_{\tilde\chi^\pm_1} >$ 103.5~GeV then only allows $\mu >0$. 
The grey shaded region in Fig.~\ref{Fig1} is excluded for not satisfying 
the electroweak symmetry breaking condition, while the brick-colored region 
is excluded because the stau is lighter than the neutralino hence neutralino 
cannot be the dark matter candidate. 
The red solid line shows the neutralino-proton elastic scattering cross-section contour 
of $9\times 10^{-9}$ pb which approximately is the bound from XENON100~\cite{xenon100} 
for 70 GeV neutralino mass. 
Note, however, that the theoretical cross-section can easily have large uncertainties due to the  
hadronic factor determinations, the dark matter profile and the galactic velocity distribution. 
We show the CDF 90\% CL contour  
of Br($B_s\to\mu^+\mu^-$), as in Eq.(\ref{90CL}), as a green shaded region.
The maximum allowed values of  $m_{1/2}$ and $m_0$ go up to $\sim 1100$~GeV for $A_0=0$ 
and $\tan\beta=40$. This covers the whole neutralino-stau coannihilation band favored by dark matter.
If we increase $\tan\beta$ to 50, the region allowed by the 90\% CL range of 
Br($B_s\to\mu^+\mu^-$) increases and the maximum allowed value of  
$m_{1/2}$, that satisfies the dark matter constraint, goes up to 1.6~TeV. 
In addition, we also show the 1$\sigma$ 
contour, Eq.(\ref{1sigma}), as dashed light-green lines in both figures.
The lowest $\tan\beta$ value, for $A_0=0$,  
allowed by the limits on the  Higgs mass and the $b\rightarrow s \gamma$ branching ratio, 
and within the CDF 1$\sigma$ region of Br($B_s\to\mu^+\mu^-$) is about 30, but  
using the 90\% CL range $\tan\beta$ can be as low as about 20.

In Table~\ref{tab:SUSYmass} we  list the upper and lower spectrums of the sparticle masses 
in the mSUGRA model for $\tan\beta$=50 and 40 for the 90\%CL and 1$\sigma$ allowed values of 
the Br($B_s\to\mu^+\mu^-$), when we satisfy the dark matter and other constraints. 

\begin{table}
\caption{Sparticle mass ranges for $\tan\beta$=50 and 40 for 90\%CL and $1\sigma$ 
allowed values of the Br($B_s\to\mu^+\mu^-$) in the dark matter allowed regions. 
The lower bound for $\tan\beta = 40$ is fixed by the $b\rightarrow s \gamma$ constraint, 
hence the same for both 90\% CL and 1$\sigma$.}
\label{tab:SUSYmass}
\begin{center}
\begin{tabular}{|c| c | c|  c| c| c| c| c| c| c| c|}
\hline \hline $\tan\beta$ &$m_{1/2}$(TeV)&$\tilde g$(GeV)&$\tilde u_1$&$\tilde d_1$&$\tilde t_1$&
$\tilde b_1$&$\tilde e_1$&$\tilde\tau_1$&$\tilde\chi^0_1$&$\tilde\chi^{\pm}_1$\\\hline
50& 0.46 & 1088 & 950 & 947 & 743 & 833 & 352 & 204 & 193 & 372 \\\cline{2-11}
&1.61 & 3418 & 2917 & 2906 & 2379 & 2646 & 1054 & 740 & 722 & 1395  \\\hline
(1$\sigma$) & 0.52 & 1211 & 1057 & 1053 & 832 & 931 & 387 & 232 & 220 & 424 \\\cline{2-11}
&0.88 & 1952 & 1674 & 1669 & 1351 & 1503 & 581 & 384 & 381 & 738  \\\hline\hline
40 & 0.36 & 879 & 758 & 756 & 588 & 680 & 246 & 155 & 150 & 287 \\\cline{2-11}
& 1.05 & 2314 & 1954 & 1944 & 1593 & 1804 & 595 & 464 & 461 & 891 \\\hline
(1$\sigma$) & 0.58 & 1341 & 1143 & 1141 & 913 & 1042 & 351 & 252 & 246 & 474 \\ \hline\hline

\end{tabular}
\end{center}
\end{table}

Because Br($B_s\to\mu^+\mu^-$) depends on the CP odd Higgs mass $m_A$
and the Higgsino mass $\mu$,
the constraints of the SUSY parameters from the CDF measurement of the branching ratio is
different in the case of the non-universal Higgs mass (NUHM) boundary condition.
The recent analysis of Br($B_s\to\mu^+\mu^-$) in NUHM can be found in
\cite{Gogoladze:2010ch}
(See also \cite{Ellis:2006jy} for earlier analysis.).

\section{Large $B_s$ mixing in SUSY GUTs}

In SUSY GUT theories,
it is often assumed that
the SUSY breaking sfermion masses
are flavor-universal,
but the off-diagonal elements of the mass matrices
 are generated by the loop effects.
The FCNC sources are the Dirac/Majorana neutrino Yukawa couplings,
which are responsible for the large neutrino mixings
\cite{Borzumati:1986qx}.
Since the left-handed leptons $(L)$ and
the right-handed down-type quarks $(D^c)$
are unified in $\bar{\bf 5}$,
the Dirac neutrino Yukawa couplings can be written as
$Y_\nu{}_{ij} \bar{\bf 5}_i N^c_j H_{\bf 5}$,
where $N^c$ is the right-handed neutrino.
The flavor non-universality of the SUSY breaking $\tilde D^c$
masses is generated
by the colored Higgs and the $N^c$ loop diagram \cite{Moroi:2000tk},
and the non-universal part of the mass matrix is
$\delta M_{\tilde D^c}^2 \simeq
- \frac1{8\pi^2} (3m_0^2+A_0^2) Y_\nu Y_\nu^\dagger \ln(M_*/M_{H_C})$,
where $M_*$ is a cut-off scale (e.g. the Planck scale), $M_{H_C}$ is
a colored Higgs mass,
$m_0$ is the universal scalar mass and
$A_0$ is the universal scalar trilinear coupling.
The left-handed Majorana neutrino coupling $LL\Delta_L$
($\Delta_L$ is an SU(2)$_L$ triplet)
can also provide  contributions to the light neutrino mass
(type II seesaw \cite{Schechter:1980gr}),
and can generate the FCNC in the sfermion masses
when the fermions are unified.

As a convention in this paper,
we will call the model with the FCNC source arising from
the Dirac neutrino Yukawa coupling as the minimal type of SU(5).
In this case, the off-diagonal elements of $\bf 10$
($Q,U^c,E^c$) representations are small because they originate from
the CKM mixings.
In a competitive model which we call the minimal type of SO(10),
the Majorana couplings, which contribute to the neutrino mass,
generate the off-diagonal elements for all sfermion species
since the Majorana couplings $f_{ij} L_iL_j\Delta_L$ can be unified to the
$f_{ij}{\bf 16}_i\ {\bf 16}_j\ \overline{\bf 126}$ coupling \cite{Babu:1992ia}.
The detail can be found in Ref.\cite{Dutta:2006gq,Dutta:2009hj}.

The NP contribution of the $B_s$-$\bar B_s$ mixing amplitude 
can be described as
\begin{equation}
M_{12}^s = M_{12,\rm SM}^s + M_{12,\rm NP}^s
=C_s \, M_{12,\rm SM}^s \, e^{2i \phi_{B_s}},
\label{parameter}
\end{equation}
where $C_s$ is a real positive number. 
From the measurement of the mass difference, $\Delta M_s = 2|M_{12}^s|$,
the experimental result is consistent with $C_s =1$.
%
%
Even if $C_s=1$,
there is room for new physics because of the phase freedom, $\phi_{B_s}$.
The phase can be measured by $B_s\to J/\psi\phi$ decay
and the dimuon asymmetry from semileptonic $B$ decays.
In the SM, the CP violation of
$B_s\to J/\psi\phi$ is tiny ($2\beta_s^{\rm SM}\sim 0.04$).
%
%
%
%
%
The CP phase $2\beta_s (= 2(\beta_s^{\rm SM}+\phi_{B_s}))$ and the decay width difference 
$\Delta \Gamma_s$ have been measured
at the Tevatron \cite{Aaltonen:2007he} and the LHCb \cite{LHCb-phis},
and a large phase is allowed.
%
%
The dimuon asymmetry reported by D0
implies a large CP phase in $B_s$-$\bar B_s$ mixing \cite{Abazov:2010hv}.

By definition in the Eq.(\ref{parameter}), we obtain
\begin{equation}
\sin^2\phi_{B_s} =
\frac{\left(\frac{A^{\rm NP}_s}{A^{\rm SM}_s}\right)^2-(1-C_s)^2}{4C_s},
\label{relation-phiBs}
\end{equation}
where $A_s^{\rm NP} = | M_{12,\rm NP}^s |$ and
$A_s^{\rm SM} = | M_{12,\rm SM}^s |$.
In the case of $C_s \simeq 1$, we obtain $2\sin\phi_{B_s} \simeq A^{\rm NP}_s/A^{\rm SM}_s$.
%
%
The large dimuon asymmetry reported by D0
requires $A^{\rm NP}_s/A^{\rm SM}_s \sim O(1)$.

In SUSY GUTs, where quark and lepton unification is manifested,
such large values of
$A^{\rm NP}_s/A^{\rm SM}_s$
indicate a large lepton flavor violation such as $\tau\to\mu\gamma$.
Therefore, in order to satisfy the current experimental bound \cite{Hayasaka:2007vc}:
\begin{equation}
{\rm Br}(\tau\to\mu\gamma) < 4.4 \times 10^{-8},
\end{equation}
the parameters in the model are constrained.
Since Br($\tau\to\mu\gamma$) is proportional to $\tan^2\beta$,
the constraint is more severe for a larger $\tan\beta$. 
The CDF reported Br($B_s\to\mu^+\mu^-$)
actually prefers a large $\tan\beta$.

In the MSSM, the $B_s$-$\bar B_s$ mixing amplitude 
is induced by the box diagram contribution
and the Higgs mediated contribution.
The box contribution does not depend on $\tan\beta$ explicitly,
while the Higgs mediated contribution is proportional to $\tan^4\beta/m_A^2$.
Therefore, for a large $\tan\beta$,
the Higgs mediated contribution can dominate over the 
box contribution,
and 
it can induce 
 a sizable value of $A^{\rm NP}_s/A^{\rm SM}_s$
satisfying the experimental bound of
Br($\tau\to\mu\gamma$) in SUSY GUTs.
The flavor violating Higgs coupling 
can induce not only a large contribution to the $B_s$-$\bar B_s$ mixing amplitude
but also to $B_s\to\mu^+\mu^-$.
As a result, 
for a given size of $A^{\rm NP}_s/A^{\rm SM}_s$,
Br($\tau\to\mu\gamma$) and Br($B_s\to\mu^+\mu^-$)
 are bounded.

\begin{figure}[tbp]
 \center
  \includegraphics[width=9cm]{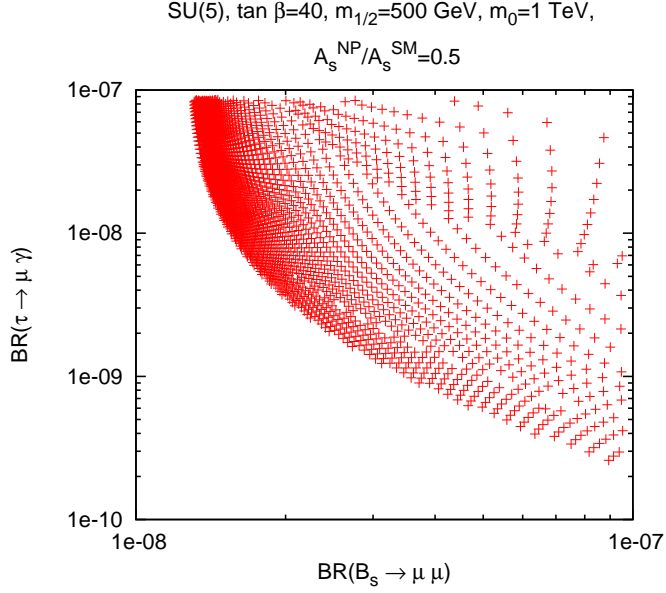}
 \caption{We plot the correlation of Br($\tau\to\mu\gamma$) and  Br($B_s\to\mu^+\mu^-$) 
for $A^{\rm NP}_s/A^{\rm SM}_s = 0.5$
in the case of the minimal type of SU(5) for universal boundary condition.
}
\label{Fig2}
\end{figure}

In Fig.\ref{Fig2},
we show the correlation of
Br($\tau\to\mu\gamma$) and Br($B_s\to\mu^+\mu^-$)
for $A^{\rm NP}_s/A^{\rm SM}_s = 0.5$
in the case of the minimal type of SU(5).
In the plot, the Higgsino mass $\mu$ and the CP odd Higgs mass $m_A$
are varied assuming that the SUSY breaking
Higgs masses $m_{H_u}$ and $m_{H_d}$ are different from the
universal sfermion mass ($m_0 = 1$ TeV).
We choose the gaugino mass $m_{1/2} = 500$ GeV
and $\tan\beta = 40$.
For a given value of Br($\tau\to\mu\gamma$), 
Br($B_s\to\mu^+\mu^-$) is bounded from below,
and vice versa.
In this example, 
in order to satisfy the CDF result
Br($B_s\to\mu^+\mu^- ) < 3.5 \times 10^{-8}$,
the branching ratio of 
$\tau\to\mu\gamma$ has to be larger than
$10^{-9}$.
Br($B_s\to\mu^+\mu^-$)
is also bounded from below
due to the Br($\tau\to\mu\gamma$) constraint.

\begin{figure}[tbp]
 \center
  \includegraphics[width=8cm]{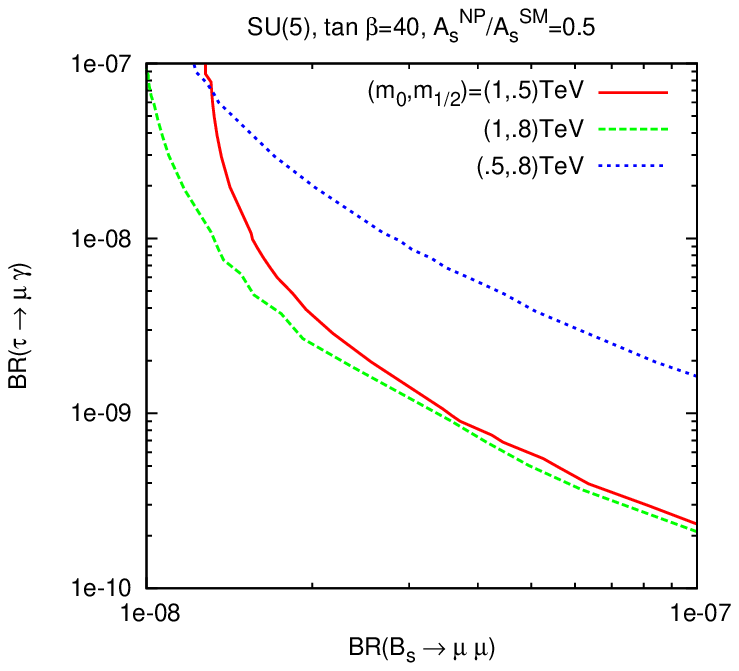}
  \includegraphics[width=8cm]{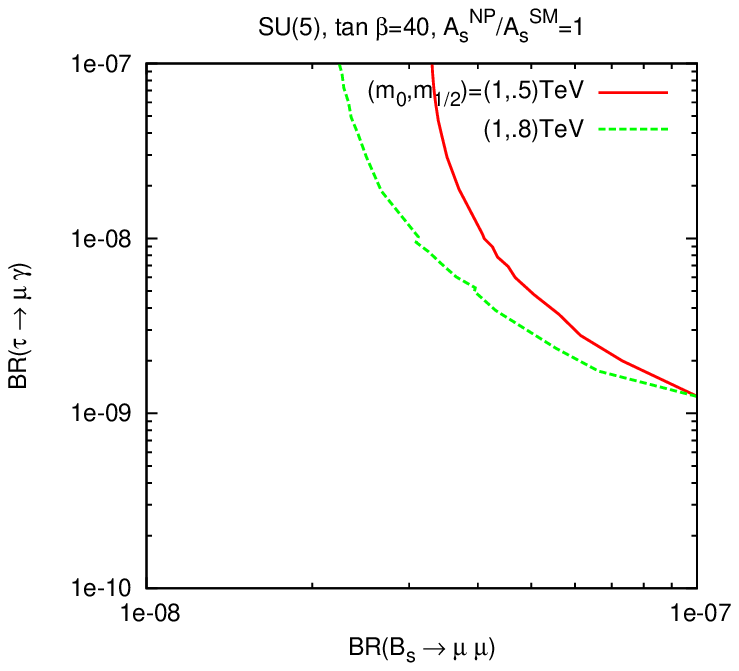}
 \caption{The boundaries
of  
Br($\tau\to\mu\gamma$) and Br($B_s\to\mu^+\mu^-$) correlation regions are shown 
for different input parameters $m_0$ and $m_{1/2}$. 
We choose
$A^{\rm NP}_s/A^{\rm SM}_s = 0.5$ (left plot) and 1 (right plot).
}
\label{Fig3}
\end{figure}

In Fig.\ref{Fig3},
we plot the boundaries
of 
Br($\tau\to\mu\gamma$) and Br($B_s\to\mu^+\mu^-$) correlation regions
in the same manner as the plot in Fig.\ref{Fig2},
while changing the input parameters $m_0$ and $m_{1/2}$.
We plot two cases $A^{\rm NP}_s/A^{\rm SM}_s = 0.5$ and 1.
Surely,
a larger $A^{\rm NP}_s/A^{\rm SM}_s = 1$
provides more stringent constraints.
The squark mass is less dependent on $m_0$
due to the gluino loop compared to the slepton mass.
Therefore, larger $m_0$ gives a smaller Br($\tau\to\mu\gamma$).
In the case of $A^{\rm NP}_s/A^{\rm SM}_s = 1$,
$m_0 = 500$ GeV is already excluded
by Br($B_s\to\mu^+\mu^-$) and Br($\tau\to\mu\gamma$) bounds.

These bounds are significant when
the quark and lepton unification is manifested.
In the minimal type of SU(5) GUT where the squark flavor 
violation is generated from Dirac neutrino Yukawa coupling,
the unification is manifested.
In the case of SO(10) GUT,
the bounds can be relaxed
by a choice of symmetry breaking vacua \cite{Dutta:2007ai}.
Therefore,
the accurate measurements of 
Br($B_s\to\mu^+\mu^-$)
and
the $B_s$-$\bar B_s$ mixing phase
are informative to distinguish
the symmetry breaking vacua.

\section{Flavor violating $\tau$ decays}

The rare decay process $B_s\to\mu^+\mu^-$ 
is dominantly generated by the Higgs mediated diagram.
The leptonic version of the Higgs mediated diagram
gives
$\tau^+\to \mu^+\mu^-\mu^+$ \cite{Babu:2002et} or
$\tau\to\mu\eta$ \cite{Sher:2002ew},
and they are also proportional to $\tan^6\beta/m_A^4$.
Therefore, the correlation between the Higgs mediated $\tau$ decay
and $B_s\to\mu^+\mu^-$
can be important to investigate the quark and lepton unification.

The current experimental bounds are 
Br($\tau\to\mu\eta) < 2.3 \times 10^{-8}$ \cite{Aubert:2006cz}
and Br($\tau\to 3\mu) < 2.1 \times 10^{-8}$ \cite{Marchiori:2009ww}.
Because the Higgs mediated contribution to these processes
generates a fixed ratio
Br($\tau\to\mu\eta$)/Br($\tau\to 3\mu$) = 8.4 \cite{Sher:2002ew},
the $\tau\to\mu\eta$ decay gives more important bounds 
to study the Higgs mediated $\tau$-$\mu$ flavor violation \cite{Dedes:2002rh}.

\begin{figure}[tbp]
 \center
  \includegraphics[width=9cm]{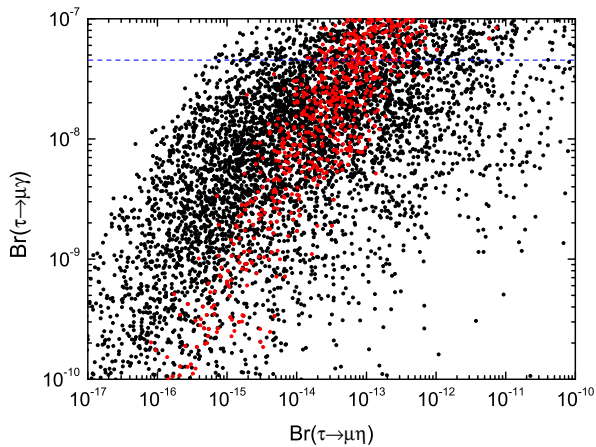}
 \caption{Br($\tau\to\mu\gamma$) is shown as a function of Br($\tau\to\mu\eta$)  
for universal boundary condition (given in the text). The red dotted points  
satisfy the CDF allowed range of Br($B_s\to\mu^+\mu^-$).  
}
\label{Fig4}
\end{figure}

The measured Br($B_s\to\mu^+\mu^-$)
has an impact on the 
Br($\tau\to\mu\eta$) for  given boundary conditions.
In Fig.\ref{Fig4},
we plot 
Br($\tau\to\mu\eta$)
and
Br($\tau\to\mu\gamma$)
for $m_{1/2} = 500$ GeV, $A_0=0$ and $\tan\beta = 50$.
We vary the SUSY breaking sfermion mass $m_0 < 1.5$ TeV.
We take the non-universal SUSY breaking Higgs masses 
to vary $\mu$ and $m_A$.
We assume that the left-handed slepton mass matrix
can have off-diagonal elements due to the right-handed neutrino loop.
Since the $\tau$-$\mu$ Higgs coupling is
induced by the finite correction, the $\tau\to\mu\gamma$ operator
is correlated and
they are proportional to each other 
for given mass spectrum.
The red points in the plot satisfy
the 1 sigma range of Br($B_s\to\mu^+\mu^-$) reported by the CDF.
The CDF result constrains
Br($\tau\to\mu\eta$)
for a given value of
Br($\tau\to\mu\gamma$).

\begin{figure}[tbp]
 \center
  \includegraphics[width=9cm]{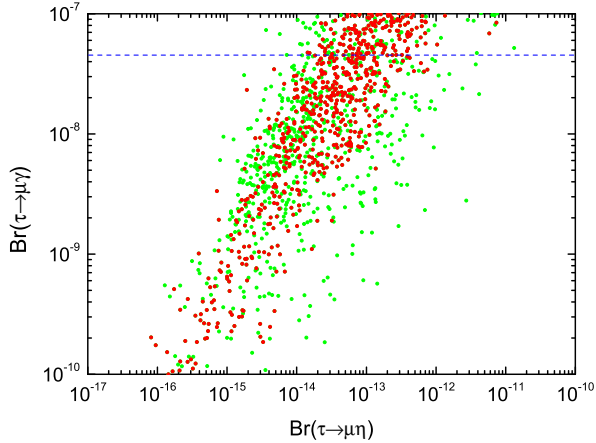}
 \caption{Br($\tau\to\mu\gamma$) vs Br($\tau\to\mu\eta$) is plotted for 
non-universal boundary condition (given in the text). The points  satisfy 
the CDF allowed range of Br($B_s\to\mu^+\mu^-$).  
The red dotted points satisfy the universal boundary conditions (as in Fig.4).
}
\label{Fig5}
\end{figure}

The Higgs mediated $b_R$-$s_L$ coupling is 
generated by stop-chargino loop diagram,
and the Higgs mediated $\tau_R$-$\mu_L$ coupling
is generated by stau-chargino loop diagram.
As a result, the 
 stop and stau mass spectrum
is important for the 
correlation between
Br($\tau\to\mu\eta$)
and 
Br($\tau\to\mu\gamma$)
for a given Br($B_s\to \mu^+\mu^-$).
Therefore, the correlation can be a probe
of the SUSY breaking sfermion mass universality and
the gaugino mass unification.
For example,
if the sfermion masses are not universal,
the correlation between
Br($\tau\to\mu\eta$)
and 
Br($\tau\to\mu\gamma$)
will be broken.
In order to illustrate it,
we show the plot for the case of sfermion mass non-universality.
In Fig.\ref{Fig5},
we plot 
Br($\tau\to\mu\eta$)
and 
Br($\tau\to\mu\gamma$)
for $m_{1/2} = 500$ GeV, $A_0=0$ and $\tan\beta=50$. 
We choose the SUSY breaking sfermion mass $m_{10} \neq m_{5}$
where 
$m_{10} = m_{\tilde Q} = m_{\tilde U} = m_{\tilde E}$
and $m_5 = m_{\tilde D} = m_{\tilde L}$.
The points satisfies the 1 sigma Br($B_s \to \mu^+\mu^-$) range
reported by CDF.
The red points correspond to the case of sfermion universality
and thus, they are same as the red points in Fig.\ref{Fig4}.
The green points correspond to the case of non-universal sfermion mass.

The correlation between
Br($\tau\to\mu\eta$)
and 
Br($\tau\to\mu\gamma$)
can be also broken if
 $b$-$s$ Higgs coupling is accidentally
canceled by a new left-handed squark flavor violating source
(which is absent in the minimal type of SU(5)).
If the sizes of the off-diagonal elements are 
same for all sfermion species,
the $\tau\to\mu\gamma$ bound can be easily saturated
and the $B_s\to \mu^+\mu^-$ amplitude will not be canceled.
Therefore, the flavor violation in the quark sector 
has to be larger than one in the lepton sector
choosing the SO(10) breaking vacua to allow the cancellation.
However,
if the $b_R$-$s_L$ Higgs coupling is canceled
(namely the new FCNC contribution is destructive), then 
the new FCNC provides constructive SUSY contribution
to the $b\to s\gamma$ operator ($C_{7L}$) \cite{Buras:2001mb},
which is disfavored from the experimental constraint.


It is expected that
the upper bounds of the branching ratios of flavor violating 
$\tau$ decays can be roughly one order below 
at the super B factory \cite{Aushev:2010bq}.
It may be hard to achieve the allowed region of Br($\tau\to\mu\eta$).
If, however, Br($\tau\to\mu\eta$) is measured to be $\agt 10^{-9}$,
the slepton and squark mass universality 
or gaugino mass unification may need to be broken to explain. 
%

\section{Conclusion}

The $b$-$s$ flavor changing transition is one of the important probes
of new physics.
The purely leptonic $B_s$ meson decays are helicity suppressed in SM,
and the prediction of the branching ratio is very small.
Therefore, 
the measurement of the rare decay process of $B_s \to \mu^+\mu^-$ 
plays an important role in finding the signature 
of the new physics beyond SM.
In fact, in the MSSM, the decay rate has a strong dependence on $\tan\beta$,
and it can be large even if the colored SUSY particles are heavy,
as preferred by the current experimental constraints.
In other words, a large $\tan\beta$ is preferred
to reproduce the large branching ratio of $B_s \to \mu^+\mu^-$
for heavy colored SUSY particles.
In this way, the CDF measurement of the 
$B_s \to \mu^+\mu^-$ rare decay process
has a tremendous impact to bound the SUSY parameters, and
can provide a crucial upper bound.

Even in the minimal models of the SUSY breaking parameters
in which flavor universality is assumed,
the $B_s\to \mu^+\mu^-$ process can be large
due to the flavor violation originated from CKM mixings.
In GUT models, the flavor violation can be related to the symmetry breaking
and the GUT particle spectrum,
and the branching ratio of 
$B_s\to \mu^+\mu^-$ process can give a constraint on the GUT models.
Especially for the minimal type of SU(5) GUT
where the flavor violation comes from the Dirac neutrino Yukawa coupling,
the parameter space is really restricted
and the branching ratios Br($B_s\to \mu^+\mu^-$) and
BR($\tau\to\mu\gamma$) are bounded from below
if the CP violation in $B_s$ mixing is large,
which is indicated by the $B_s \to J/\psi\phi$ decay reported by the Tevatron and the LHCb,
and the dimuon asymmetry in the semileptonic $B$ decays reported by D0.
The flavor violating lepton decays are also important
to find a footprint of the GUT scale physics.

The excess of the $B_s\to\mu^+\mu^-$ decay deviation from the SM prediction
 at more than 90\% CL
can be soon verified at the LHC,
and the squarks and the gluino can be soon found
if the true value of Br($B_s\to\mu^+\mu^-$) lies in the 1 sigma range of CDF measurements.
The large $B_s$ CP phase can be also soon verified at the LHC.
A surge of  experimental results
will appear soon 
to test the whole structure of the models presented.

\section*{Acknowledgments}
The work of B.D. is supported in part by the DOE grant DE-FG02-95ER40917.
The work of Y.M. is supported by the Excellent Research Projects of
National Taiwan University under grant number NTU-98R0526. 
We thank Teruki Kamon for various discussions.


\begin{thebibliography}{99}
%
%
%
%
%
%
%
%



\bibitem{CDF}
  CDF~Collaboration,
  arXiv:1107.2304 [hep-ex].


\bibitem{Buchalla:1993bv}
  G.~Buchalla and A.~J.~Buras,
  Nucl.\ Phys.\  B {\bf 400}, 225 (1993);
%
  A.~J.~Buras,
  Phys.\ Lett.\  B {\bf 566}, 115 (2003)
  [hep-ph/0303060].


\bibitem{Gamiz:2009ku}
  E.~Gamiz, C.~T.~H.~Davies, G.~P.~Lepage, J.~Shigemitsu and M.~Wingate
                  [HPQCD Collaboration],
  Phys.\ Rev.\  D {\bf 80}, 014503 (2009)
  [arXiv:0902.1815 [hep-lat]].



\bibitem{Aaij:2011rj}
  R.~Aaij {\it et al.}  [the LHCb Collaboration],
  Phys.\ Lett.\  B {\bf 699}, 330 (2011)
  [arXiv:1103.2465 [hep-ex]].



\bibitem{Choudhury:1998ze}
  S.~R.~Choudhury and N.~Gaur,
  Phys.\ Lett.\  B {\bf 451}, 86 (1999)
  [hep-ph/9810307];
%
  K.~S.~Babu and C.~F.~Kolda,
  Phys.\ Rev.\ Lett.\  {\bf 84}, 228 (2000)
  [hep-ph/9909476];
%
  C.~S.~Huang, W.~Liao, Q.~S.~Yan and S.~H.~Zhu,
  Phys.\ Rev.\  D {\bf 63}, 114021 (2001)
  [Erratum-ibid.\  D {\bf 64}, 059902 (2001)]
  [hep-ph/0006250];
%
  P.~H.~Chankowski and L.~Slawianowska,
  Phys.\ Rev.\  D {\bf 63}, 054012 (2001)
  [hep-ph/0008046];
%
  C.~Bobeth, T.~Ewerth, F.~Kruger and J.~Urban,
  Phys.\ Rev.\  D {\bf 64}, 074014 (2001)
  [hep-ph/0104284];
%
  Phys.\ Rev.\  D {\bf 66}, 074021 (2002)
  [hep-ph/0204225].


\bibitem{Borzumati:1986qx}
  F.~Borzumati and A.~Masiero,
  Phys.\ Rev.\ Lett.\  {\bf 57}, 961 (1986);
%
  J.~Hisano, 
  T.~Moroi, K.~Tobe, M.~Yamaguchi and T.~Yanagida,
  Phys.\ Lett.\ B {\bf 357}, 579 (1995)
  [hep-ph/9501407].


\bibitem{Barbieri:1994pv}
  L.~J.~Hall, V.~A.~Kostelecky and S.~Raby,
  Nucl.\ Phys.\ B {\bf 267}, 415 (1986);
%
  R.~Barbieri and L.~J.~Hall,
  Phys.\ Lett.\ B {\bf 338}, 212 (1994)
  [hep-ph/9408406];
%
  J.~Hisano, T.~Moroi, K.~Tobe and M.~Yamaguchi,
  Phys.\ Lett.\ B {\bf 391}, 341 (1997)
  [hep-ph/9605296].



\bibitem{Moroi:2000tk}
  T.~Moroi,
  Phys.\ Lett.\  B {\bf 493}, 366 (2000)
  [hep-ph/0007328];
%
D.~Chang, A.~Masiero and H.~Murayama,
  Phys.\ Rev.\  D {\bf 67}, 075013 (2003)
  [hep-ph/0205111].
%
R.~Harnik, D.~T.~Larson, H.~Murayama and A.~Pierce,
  Phys.\ Rev.\  D {\bf 69}, 094024 (2004)
  [hep-ph/0212180].



\bibitem{Dutta:2007ai}
  B.~Dutta, Y.~Mimura and R.~N.~Mohapatra,
  Phys.\ Rev.\ Lett.\  {\bf 100}, 181801 (2008)
  [arXiv:0712.1206 [hep-ph]].






\bibitem{Parry:2005fp}
  J.~K.~Parry,
  Nucl.\ Phys.\  B {\bf 760}, 38 (2007)
  [hep-ph/0510305];
%
  Mod.\ Phys.\ Lett.\  A {\bf 21}, 2853 (2006)
  [hep-ph/0608192];
%
  J.~K.~Parry and H.~h.~Zhang,
  Nucl.\ Phys.\  B {\bf 802}, 63 (2008)
  [arXiv:0710.5443 [hep-ph]];
%
  J.~K.~Parry,
  Phys.\ Lett.\  B {\bf 694}, 363 (2011)
  [arXiv:1006.5331 [hep-ph]].

\bibitem{Dutta:2006gq}
  B.~Dutta and Y.~Mimura,
  Phys.\ Rev.\ Lett.\  {\bf 97}, 241802 (2006)
  [hep-ph/0607147];
%
  Phys.\ Rev.\  D {\bf 75}, 015006 (2007)
  [hep-ph/0611268];
%
  Phys.\ Rev.\  D {\bf 77}, 051701 (2008)
  [arXiv:0708.3080 [hep-ph]];
%
  Phys.\ Rev.\  D {\bf 78}, 071702 (2008)
  [arXiv:0805.2988 [hep-ph]].

\bibitem{Goto:2007ee}
  T.~Goto, Y.~Okada, T.~Shindou and M.~Tanaka,
  Phys.\ Rev.\  D {\bf 77}, 095010 (2008)
  [arXiv:0711.2935 [hep-ph]].


\bibitem{Hisano:2008df}
  J.~Hisano and Y.~Shimizu,
  Phys.\ Lett.\  B {\bf 669}, 301 (2008)
  [arXiv:0805.3327 [hep-ph]];
%
  J.~h.~Park and M.~Yamaguchi,
  Phys.\ Lett.\  B {\bf 670}, 356 (2009)
  [arXiv:0809.2614 [hep-ph]].



\bibitem{Hamzaoui:1998nu}
  C.~Hamzaoui, M.~Pospelov and M.~Toharia,
  Phys.\ Rev.\  D {\bf 59}, 095005 (1999)
  [hep-ph/9807350];
%
  M.~Gorbahn, S.~Jager, U.~Nierste and S.~Trine,
  arXiv:0901.2065 [hep-ph].


\bibitem{Buras:2001mb}
  A.~J.~Buras, P.~H.~Chankowski, J.~Rosiek and L.~Slawianowska,
  Nucl.\ Phys.\  B {\bf 619}, 434 (2001)
  [hep-ph/0107048];
%
  Phys.\ Lett.\  B {\bf 546}, 96 (2002)
  [hep-ph/0207241];
%
  Nucl.\ Phys.\  B {\bf 659}, 3 (2003)
  [hep-ph/0210145].






\bibitem{Foster:2004vp}
  J.~Foster, K.~Okumura and L.~Roszkowski,
  Phys.\ Lett.\  B {\bf 609}, 102 (2005)
  [hep-ph/0410323];
%
  JHEP {\bf 0508}, 094 (2005)
  [hep-ph/0506146];
%
  Phys.\ Lett.\  B {\bf 641}, 452 (2006)
  [hep-ph/0604121].




\bibitem{Dutta:2009iy}
  B.~Dutta and Y.~Mimura,
  Phys.\ Lett.\  B {\bf 677}, 164 (2009)
  [arXiv:0902.0016 [hep-ph]].


\bibitem{Aaltonen:2007he}
  T.~Aaltonen {\it et al.}  [CDF Collaboration],
  Phys.\ Rev.\ Lett.\  {\bf 100}, 161802 (2008)
  [arXiv:0712.2397 [hep-ex]];
%
  V.~M.~Abazov {\it et al.}  [D0 Collaboration],
  Phys.\ Rev.\ Lett.\  {\bf 101}, 241801 (2008)
  [arXiv:0802.2255 [hep-ex]].
%

\bibitem{LHCb-phis}
The LHCb Collaboration, LHCb-CONF-2011-006.

\bibitem{Abazov:2010hv}
  V.~M.~Abazov {\it et al.}  [D0 Collaboration],
  Phys.\ Rev.\  D {\bf 82}, 032001 (2010)
  [arXiv:1005.2757 [hep-ex]];
%
  Phys.\ Rev.\ Lett.\  {\bf 105}, 081801 (2010)
  [arXiv:1007.0395 [hep-ex]];
%
  arXiv:1106.6308 [hep-ex].


\bibitem{Dutta:2009hj}
  B.~Dutta, Y.~Mimura and Y.~Santoso,
  Phys.\ Rev.\  D {\bf 80}, 095005 (2009)
  [arXiv:0907.4946 [hep-ph]];
%
  Phys.\ Rev.\  D {\bf 82}, 055017 (2010)
  [arXiv:1007.3696 [hep-ph]].

\bibitem{msugra}
  A.~H.~Chamseddine, R.~L.~Arnowitt, P.~Nath,
  Phys.\ Rev.\ Lett.\  {\bf 49}, 970 (1982);
  R.~Barbieri, S.~Ferrara, C.~A.~Savoy,
  Phys.\ Lett.\  {\bf B119}, 343 (1982);
  L.~J.~Hall, J.~D.~Lykken, S.~Weinberg,
  Phys.\ Rev.\  {\bf D27}, 2359-2378 (1983);
  P.~Nath, R.~L.~Arnowitt, A.~H.~Chamseddine,
  Nucl.\ Phys.\  {\bf B227}, 121 (1983);
 For a review, see
  H.~P.~Nilles,
  Phys.\ Rept.\  {\bf 110}, 1-162 (1984).
  

\bibitem{Arnowitt:2002cq}
  R.~L.~Arnowitt, B.~Dutta, T.~Kamon and M.~Tanaka,
  Phys.\ Lett.\  B {\bf 538}, 121 (2002)
  [hep-ph/0203069];
  A.~Dedes, H.~K.~Dreiner, U.~Nierste,
  Phys.\ Rev.\ Lett.\  {\bf 87}, 251804 (2001).
  [hep-ph/0108037];
  H.~Baer, C.~Balazs, A.~Belyaev, J.~K.~Mizukoshi, X.~Tata, Y.~Wang,
  JHEP {\bf 0207}, 050 (2002).
  [hep-ph/0205325];
  J.~R.~Ellis, K.~A.~Olive, V.~C.~Spanos,
  Phys.\ Lett.\  {\bf B624}, 47-59 (2005).
  [hep-ph/0504196].

\bibitem{ppc}Talk presented at PPC 2011, CERN, by Beate Hein, ATLAS-conf-2011-086.


\bibitem{higgs1}
ALEPH, DELPHI, L3, OPAL Collaborations,
G. Abbiendi, {\it et al.},
(The LEP Working Group for Higgs Boson Searches),
 Phys.\ Lett.\ B {\bf 565}, 61 (2003).


\bibitem{FeynHiggs}
  T.~Hahn, S.~Heinemeyer, W.~Hollik, H.~Rzehak and G.~Weiglein,
  Comput.\ Phys.\ Commun.\  {\bf 180} (2009) 1426.


\bibitem{bsgamma}
S.~Chen {\it et al.}  [CLEO Collaboration],
Phys.\ Rev.\ Lett.\  {\bf 87} (2001) 251807
[hep-ex/0108032];
%
P.~Koppenburg {\it et al.}  [Belle Collaboration],
Phys.\ Rev.\ Lett.\  {\bf 93} (2004) 061803
[hep-ex/0403004];
%
B.~Aubert {\it et al.}  [BaBar Collaboration],
hep-ex/0207076;
%
 M.~Ciuchini, G.~Degrassi, P.~Gambino and G.~F.~Giudice,
  Nucl.\ Phys.\ B {\bf 527} (1998) 21
  [hep-ph/9710335];
%
  Nucl.\ Phys.\ B {\bf 534} (1998) 3
  [hep-ph/9806308];
%
  C. Degrassi, P. Gambino and G.~F. Giudice,
JHEP {\bf 0012} (2000) 009 [hep-ph/0009337];
P.~Gambino and M.~Misiak,
Nucl.\ Phys.\ B {\bf 611} (2001) 338.



\bibitem{wmap} 
E.~Komatsu {\it et al.} [WMAP Collaboration],
  Astrophys.\ J.\ Suppl.\  {\bf 192}, 18 (2011).
  [arXiv:1001.4538 [astro-ph.CO]].



\bibitem{aleph}
  K.~Nakamura {\it et al.}  [Particle Data Group],
  J.\ Phys.\ G {\bf 37}, 075021 (2010).


\bibitem{amu}
Muon $g-2$ Collaboration, G. Bennett {\it et al.},
Phys.\ Rev.\ Lett.\ {\bf 74}, 161802 (2004).
%
%




\bibitem{Teubner:2010ah}
  T.~Teubner, K.~Hagiwara, R.~Liao, A.~D.~Martin and D.~Nomura,
  arXiv:1001.5401 [hep-ph].


\bibitem{Davier:2009zi}
  M.~Davier, A.~Hoecker, B.~Malaescu, C.~Z.~Yuan and Z.~Zhang,
  Eur.\ Phys.\ J.\  C {\bf 66}, 1 (2010)
  [arXiv:0908.4300 [hep-ph]].









  
\bibitem{xenon100}
E.~Aprile {\it et al.} [XENON100 Collaboration],
  arXiv:1104.2549 [astro-ph.CO].



\bibitem{Gogoladze:2010ch}
  I.~Gogoladze, R.~Khalid, Y.~Mimura and Q.~Shafi,
  Phys.\ Rev.\  D {\bf 83}, 095007 (2011)
  [arXiv:1012.1613 [hep-ph]].

\bibitem{Ellis:2006jy}
  J.~R.~Ellis, K.~A.~Olive, Y.~Santoso and V.~C.~Spanos,
  JHEP {\bf 0605}, 063 (2006)
  [arXiv:hep-ph/0603136].








\bibitem{Schechter:1980gr}
  J.~Schechter and J.~W.~F.~Valle,
  Phys.\ Rev.\ D {\bf 22}, 2227 (1980);
%
  R.~N.~Mohapatra and G.~Senjanovic,
  Phys.\ Rev.\ D {\bf 23}, 165 (1981);
%
  G.~Lazarides, Q.~Shafi and C.~Wetterich,
  Nucl.\ Phys.\ B {\bf 181}, 287 (1981).



\bibitem{Babu:1992ia}
  K.~S.~Babu and R.~N.~Mohapatra,
  Phys.\ Rev.\ Lett.\  {\bf 70}, 2845 (1993)
  [hep-ph/9209215].



\bibitem{Hayasaka:2007vc}
  K.~Hayasaka {\it et al.}  [Belle Collaboration],
  Phys.\ Lett.\  B {\bf 666}, 16 (2008)
  [arXiv:0705.0650 [hep-ex]];
%
  B.~Aubert {\it et al.}  [BABAR Collaboration],
  Phys.\ Rev.\ Lett.\  {\bf 104}, 021802 (2010)
  [arXiv:0908.2381 [hep-ex]].







\bibitem{Babu:2002et}
  K.~S.~Babu and C.~Kolda,
  Phys.\ Rev.\ Lett.\  {\bf 89}, 241802 (2002)
  [hep-ph/0206310].





\bibitem{Sher:2002ew}
  M.~Sher,
  Phys.\ Rev.\  D {\bf 66}, 057301 (2002)
  [hep-ph/0207136].






\bibitem{Aubert:2006cz}
  B.~Aubert {\it et al.}  [BABAR Collaboration],
  Phys.\ Rev.\ Lett.\  {\bf 98}, 061803 (2007)
  [hep-ex/0610067];
%
  Y.~Miyazaki {\it et al.}  [BELLE Collaboration],
  Phys.\ Lett.\  B {\bf 648}, 341 (2007)
  [hep-ex/0703009];
%
  K.~Hayasaka, 
  arXiv:1103.0094 [hep-ex].



\bibitem{Marchiori:2009ww}
  G.~Marchiori  [BABAR Collaboration],
  AIP Conf.\ Proc.\  {\bf 1200}, 857 (2010)
  [arXiv:0909.3870 [hep-ex]];
%
  K.~Hayasaka {\it et al.} [Belle Collaboration],
  Phys.\ Lett.\  B {\bf 687}, 139 (2010)
  [arXiv:1001.3221 [hep-ex]].



\bibitem{Dedes:2002rh}
  A.~Dedes, J.~R.~Ellis and M.~Raidal,
  Phys.\ Lett.\  B {\bf 549}, 159 (2002)
  [hep-ph/0209207];
%
  A.~Brignole and A.~Rossi,
  Nucl.\ Phys.\  B {\bf 701}, 3 (2004)
  [hep-ph/0404211];
%
  S.~Kanemura, T.~Ota and K.~Tsumura,
  Phys.\ Rev.\  D {\bf 73}, 016006 (2006)
  [hep-ph/0505191];
%
  P.~Paradisi,
  JHEP {\bf 0602}, 050 (2006)
  [hep-ph/0508054];
%
  E.~Arganda, M.~J.~Herrero and J.~Portoles,
  JHEP {\bf 0806}, 079 (2008)
  [arXiv:0803.2039 [hep-ph]];
%
  M.~J.~Herrero, J.~Portoles and A.~M.~Rodriguez-Sanchez,
  Phys.\ Rev.\  D {\bf 80}, 015023 (2009)
  [arXiv:0903.5151 [hep-ph]];
%
  M.~Cannoni and O.~Panella,
  Phys.\ Rev.\  D {\bf 81}, 036009 (2010)
  [arXiv:0910.3316 [hep-ph]].



\bibitem{Aushev:2010bq}
  T.~Aushev {\it et al.},
  arXiv:1002.5012 [hep-ex].




\end{thebibliography}
\end{document}